\documentclass[twocolumn]{bmcart}% uncomment this for twocolumn layout and comment line below
\usepackage{amssymb,amsmath}
\usepackage[dvips]{graphicx}
\usepackage[utf8]{inputenc} %unicode support
\startlocaldefs
\endlocaldefs

\newcommand{\fdr}{F_{\text{D}}}
\newcommand{\fl}{F_{\text{L}}}
\newcommand{\vecfl}{\vec F_{\text{L}}}
\newcommand{\cd}{C_{\text{D}}}
\newcommand{\cl}{C_{\text{L}}}
\newcommand{\rd}{r_{\text{d}}}
\newcommand{\rdone}{r_{\text{d},1}}
\newcommand{\rhog}{\rho_{\text{g}}}
\newcommand{\omegad}{\omega_{\text{d}}}
\newcommand{\Omegak}{\Omega_{\text{K}}}
\newcommand{\vecomegad}{\vec \omega_{\text{d}}}

\newcommand{\vdd}{v_{\text{d-d}}}
\newcommand{\gd}{g_{\text{D}}}
\newcommand{\gl}{g_{\text{L}}}
\newcommand{\glm}{g_{\text{L,max}}}

\newcommand{\wrel}{w_{\text{rel}}}
\newcommand{\rhoint}{\rho_{\text{int}}}
\newcommand{\st}{\text{St}}
\newcommand{\tsd}{t_\text{spindown}}
\newcommand{\tcol}{t_\text{col}}
\newcommand{\ts}{t_\text{s}}
\newcommand{\Sigmag}{\Sigma_\text{g}}
\newcommand{\Sigmad}{\Sigma_\text{d}}
\newcommand{\cs}{c_\text{s}}
\newcommand{\cso}{c_\text{s,0}}
\newcommand{\hg}{H_\text{g}}
\newcommand{\hd}{H_\text{d}}
\newcommand{\nd}{n_\text{d}}
\newcommand{\fsd}{f_\text{d}}
\newcommand{\mfp}{\lambda_\text{mfp}}
\newcommand{\md}{m_\text{d}}
\newcommand{\rld}{R_\text{LD}}
\newcommand{\sigmamol}{\sigma_\text{mol}}
\newcommand{\id}{I_\text{d}}
\newcommand{\nuturb}{\nu_\text{turb}}

\newcommand{\vk}{v_{\rm K}}
\newcommand{\rdmax}{r_{\rm d,max}}
\newcommand{\rdmin}{r_{\rm d,min}}

\begin{document}

%%% Start of article front matter
\begin{frontmatter}

\begin{fmbox}
\dochead{Research}

%%%%%%%%%%%%%%%%%%%%%%%%%%%%%%%%%%%%%%%%%%%%%%
%%                                          %%
%% Enter the title of your article here     %%
%%                                          %%
%%%%%%%%%%%%%%%%%%%%%%%%%%%%%%%%%%%%%%%%%%%%%%

\title{Effect of lift force on the aerodynamics of
dust grains in the protoplanetary disk}

%%%%%%%%%%%%%%%%%%%%%%%%%%%%%%%%%%%%%%%%%%%%%%
%%                                          %%
%% Enter the authors here                   %%
%%                                          %%
%% Specify information, if available,       %%
%% in the form:                             %%
%%   <key>={<id1>,<id2>}                    %%
%%   <key>=                                 %%
%% Comment or delete the keys which are     %%
%% not used. Repeat \author command as much %%
%% as required.                             %%
%%                                          %%
%%%%%%%%%%%%%%%%%%%%%%%%%%%%%%%%%%%%%%%%%%%%%%

\author[
   addressref={aff1},                   % id's of addresses, e.g. {aff1,aff2}
   corref={aff1},                       % id of corresponding address, if any
 %  noteref={n1},                        % id's of article notes, if any
   email={masaki.yamaguchi@nao.ac.jp}   % email address
]{\inits{MSY}\fnm{Masaki S} \snm{Yamaguchi}}
\author[
   addressref={aff2},
   email={kimura@vega.ess.sci.osaka-u.ac.jp}
]{\inits{SSK}\fnm{Shigeo S} \snm{Kimura}}
%   email={jane.e.doe@cambridge.co.uk}   % email address
%]{\inits{JE}\fnm{Jane E} \snm{Doe}}
%\author[
%   addressref={aff1,aff2},
%   email={john.RS.Smith@cambridge.co.uk}
%]{\inits{JRS}\fnm{John RS} \snm{Smith}}

%%%%%%%%%%%%%%%%%%%%%%%%%%%%%%%%%%%%%%%%%%%%%%
%%                                          %%
%% Enter the authors' addresses here        %%
%%                                          %%
%% Repeat \address commands as much as      %%
%% required.                                %%
%%                                          %%
%%%%%%%%%%%%%%%%%%%%%%%%%%%%%%%%%%%%%%%%%%%%%%

\address[id=aff1]{%                           % unique id
%  \orgname{Department of Zoology, Cambridge}, % university, etc
%  \street{Waterloo Road},                     %
%  %\postcode{}                                % post or zip code
%  \city{London},                              % city
%  \cny{UK}                                    % country
%}
%\address[id=aff2]{%
%  \orgname{Marine Ecology Department, Institute of Marine Sciences Kiel},
%  \street{D\"{u}sternbrooker Weg 20},
%  \postcode{24105}
%  \city{Kiel},
%  \cny{Germany}
%}
  \orgname{National Astronomical Observatory}, % university, etc
  \street{2-21-1 Osawa, Mitaka},                     %
  %\postcode{}                                % post or zip code
  \city{Tokyo},                              % city
  \cny{Japan}                                    % country
}
\address[id=aff2]{%
  \orgname{Department of Earth and Space Science,
Graduate School of Science,
Osaka University},
  \street{1-1 Machikaneyama, Toyonaka},
 % \postcode{24105}
  \city{Osaka},
  \cny{Japan}
}
%
%%%%%%%%%%%%%%%%%%%%%%%%%%%%%%%%%%%%%%%%%%%%%%
%%                                          %%
%% Enter short notes here                   %%
%%                                          %%
%% Short notes will be after addresses      %%
%% on first page.                           %%
%%                                          %%
%%%%%%%%%%%%%%%%%%%%%%%%%%%%%%%%%%%%%%%%%%%%%%

%\begin{artnotes}
%%\note{Sample of title note}     % note to the article
%%\note[id=n1]{Equal contributor} % note, connected to author
%\end{artnotes}

\end{fmbox}% comment this for two column layout

%%%%%%%%%%%%%%%%%%%%%%%%%%%%%%%%%%%%%%%%%%%%%%
%%                                          %%
%% The Abstract begins here                 %%
%%                                          %%
%% Please refer to the Instructions for     %%
%% authors on http://www.biomedcentral.com  %%
%% and include the section headings         %%
%% accordingly for your article type.       %%
%%                                          %%
%%%%%%%%%%%%%%%%%%%%%%%%%%%%%%%%%%%%%%%%%%%%%%

\begin{abstractbox}

\begin{abstract} % abstract
We newly introduce {\it lift force} into the aerodynamics of dust grains
in the protoplanetary disk.
Although many authors have so far investigated the effects
of the drag force,
gravitational force and electric force on the dust grains,
the lift force has never been considered as a force exerted
on the dust grains in the gas disk.
If the grains are spinning and moving in the fluid, then
the lift force is exerted on them.
We show in this paper
that the dust grains can be continuously spinning due to the frequent
collisions so that the lift force continues to be exerted on them,
which is valid in a certain parameter space where the grain size
is larger than $\sim 1$ m and where the distance from the central star
is larger than 1 AU for the minimum mass solar nebula.
In addition, we estimate the effects of the force on the grain motion
and obtain the result that 
the mean relative velocity between the grains due to the lift force
is comparable to the gas velocity in the Kepler rotational frame
when the Stokes number and lift-drag ratio are both $\sim 1$.
This estimation is performed under the assumptions of the steady state
and the isotropic spin angular momentum.
We also estimate the mean relative velocity when the grains keep
spinning and conclude that the lift force marginally affects
the mean relative velocity in the minimum mass solar nebula.
If there is a grain-concentrated part in the disk, the relative velocity due to
the lift force may dominate there because of high collision rate.
%\parttitle{First part title} %if any
%Text for this section.
%
%\parttitle{Second part title} %if any
%Text for this section.
\end{abstract}

%%%%%%%%%%%%%%%%%%%%%%%%%%%%%%%%%%%%%%%%%%%%%%
%%                                          %%
%% The keywords begin here                  %%
%%                                          %%
%% Put each keyword in separate \kwd{}.     %%
%%                                          %%
%%%%%%%%%%%%%%%%%%%%%%%%%%%%%%%%%%%%%%%%%%%%%%

\begin{keyword}
\kwd{gas disk: protoplanetary}
\kwd{aerodynamics: dust grains}
%\kwd{sample}
%\kwd{article}
%\kwd{author}
\end{keyword}

% MSC classifications codes, if any
%\begin{keyword}[class=AMS]
%\kwd[Primary ]{}
%\kwd{}
%\kwd[; secondary ]{}
%\end{keyword}

\end{abstractbox}
%
%\end{fmbox}% uncomment this for twcolumn layout

\end{frontmatter}

%%%%%%%%%%%%%%%%%%%%%%%%%%%%%%%%%%%%%%%%%%%%%%
%%                                          %%
%% The Main Body begins here                %%
%%                                          %%
%% Please refer to the instructions for     %%
%% authors on:                              %%
%% http://www.biomedcentral.com/info/authors%%
%% and include the section headings         %%
%% accordingly for your article type.       %%
%%                                          %%
%% See the Results and Discussion section   %%
%% for details on how to create sub-sections%%
%%                                          %%
%% use \cite{...} to cite references        %%
%%  \cite{koon} and                         %%
%%  \cite{oreg,khar,zvai,xjon,schn,pond}    %%
%%  \nocite{smith,marg,hunn,advi,koha,mouse}%%
%%                                          %%
%%%%%%%%%%%%%%%%%%%%%%%%%%%%%%%%%%%%%%%%%%%%%%

%%%%%%%%%%%%%%%%%%%%%%%%% start of article main body
% <put your article body there>

%%%%%%%%%%%%%%%%
%% Background %%
%%
\section{Background}
\label{intro}
In the theory of the planet formation, the issue concerning the radial
drift of the meter-size dust remains an open question.
In the typical scenario $\mu$m-size dust grains grow up to be km-size
planetesimals via the collision and merging in the protoplanetary disk
\cite{gol73,hay81}.
When the dust grain grows to be meter-sized, it has a velocity
with respect to the disk gas to lose its angular momentum due to
the drag force.
Thus, the grain falls down to the central star, so that
it cannot grow further
\cite{ada76,bra08}.

Various scenarios are proposed for solving the issue on the meter-size dust.
The gravitational instability in the dust layer was investigated at first
\cite{gol73,sek83}.
In this scenario the dust grains settle toward the mid-plane to form
the dense layer, which then fragments into precursors of the planetesimals.
However, the sedimentation of the grains leads to the vertical shear
of the rotational velocity in the dust layer,
which causes the turbulence due to Kelvin-Helmholtz instability.
As a result, the grains cannot settle enough to form planetesimals
\cite{sek98}.
On the other hand, the effects of turbulence due to magneto-rotational
instability were considered \cite{bal91,san00}.
The collision between dust grains occurs more frequently by the increase
of the relative velocity due to the turbulence, so that
the growth rate of the grains can increase.
\cite{bra08}, \cite{kat13}, and \cite{oku12} have taken into account these effects of
the turbulence, and while the first one has found that the dust grains
fall down to the central star if the grain density is relatively
large, the second and third ones have found that the dust grains can grow 
sufficiently rapidly to avoid the issue of the meter-size dust when the grains are
fluffy. % \cite{kat13}
As another scenario \cite{you05} has suggested, the planetesimals
are formed by the streaming instability caused by the interaction between
the dust grains and disk gas.

We newly introduce {\it lift force} as a factor affecting the relative velocity
between the dust grains.
When a grain moves in fluid and when the fluid around the grain has
a circulation, the lift force is exerted on the grain
perpendicularly to the grain velocity, generally represented as,
\begin{equation}
\fl =\cl \cdot \frac{\pi \rd^2}{\md}\cdot \frac{1}{2}\rhog |\vec u|^2\ , 
\end{equation}
where $\cl$ and $\vec u$ is a coefficient of the lift force
and a velocity of a dust grain relative to the disk gas, respectively.
We note here that $\fl$ is defined as the lift force per unit mass.
The coefficient of the lift force is determined by properties of
the grain and flow.
For a rotating sphere, the lift force is expressed as
\begin{equation}
\vecfl =\frac{\pi\rhog \rd^3}{\md}\vecomegad \times \vec u,
\label{fl}
\end{equation}
where the Stokes law, which is valid when the Reynolds number is small,
is adopted as the drag law \cite{rub61,tak74}.
In this case, the coefficient of the lift force is represented as
\begin{equation}
\cl =\frac{2\md\fl}{\pi \rd^2 \rhog u^2}=\frac{2\rd\omegad}{u}\sin\theta,
\end{equation}
where $\theta$ is the angle between $\vecomegad$ and $\vec u$.
When the Reynolds number is so large that the turbulent flow is dominated,
and when the Knudsen number is so large that the fluid cannot be regarded
as a continuum, the lift force has not yet been formulated.
Therefore, in this paper, we investigate the effects of the lift force  
only when Equation (\ref{fl}) can be applied.
We estimate the conditions under which the lift force is kept exerted on the grain in
Section \ref{sustainability}.
We derive and reduce an equation of motion for the grain to estimate the relative velocity
between the grains in Section \ref{velo}.
In Section \ref{dis}, we evaluate the relative velocity between the grains
when the lift force is kept exerted on the grains and
discuss improvement of our model.
Finally we summarize our study in Section \ref{sum}.

\section{Sustainability of the spin of the dust grains}\label{sustainability}
In this section, we examine whether the dust grains keep spinning in the gas disk 
because the lift force does not act on the non-spinning spherical object. 
Here, we assume that the collisions between the dust grains induce the spin of the dust grains.
The spinning dust is subjected to the torque due to the friction by the background viscous fluid. 
After the spin-down time scale, the spin of the dust would stop. 
We estimate the collision time $\tcol$ and the spin-down time $\tsd$. 
By comparing these timescales, we obtain the parameter space where the lift force can act on the spinning dust grains. 
These timescales depend on the disk structure. 
We adopt the parameters for the disk structure in this paper as follows:

% \begin{eqnarray}
% \Sigmag=1.7\times10^3 R_0^{-3/2}\  \rm g\  cm^{-2},\\
% \Sigmad=0.01 \Sigmag=17 R_0^{-3/2}\  \rm g\ cm^{-2}, \label{eq:sigmad}\\
% \cs=1.2\times10^5R_0^{-1/4}\  \rm cm\  s^{-1},\\
% \Omegak=\sqrt{\frac{GM_{\text{s}}}{R^3}}=2.0\times10^{-7} R_0^{-3/2}\  \rm sec^{-1},\\
% \hg=\frac{\cs}{\Omegak}=6.0\times10^{11} R_0^{5/4}\ \rm cm, \label{eq:hg}\\
% \rhog=1.4\times 10^{-9} R_0^{-11/4}\  \rm g\ cm^{-3}, \label{eq:rhog}
% \end{eqnarray}
\begin{eqnarray}
\Sigmag=\Sigma_0 R_1^{-q},\\
\Sigmad=\fsd \Sigmag=\fsd\Sigma_0 R_1^{-q}, \label{eq:sigmad}\\
\cs =\sqrt{\frac{k T}{\overline m}}=\cso R_1^{-p} ,\\
\Omegak=\sqrt{\frac{GM_{\text{s}}}{R^3}}=\Omega_0 R_1^{-3/2},\\
\vk=R\Omegak=\Omega_0 R_0 R_1^{-1/2},\\
\hg=\frac{\sqrt{2}\cs}{\Omegak}=\frac{\sqrt{2}\cso}{\Omega_0}R_1^{-p+3/2}, \label{eq:hg}\\
\rhog= \frac{\Sigmag}{\sqrt{\pi}\hg}=\frac{\Sigma_0\Omega_0}{\sqrt{2\pi}\cso}R_1^{p-q-3/2}, \label{eq:rhog}
\end{eqnarray}
where $R$ is the semi-major axis, 
$R_0$ is typical radius of the disk, 
$R_1 = R/R_0$, 
$\fsd$ is the dust-to-gas mass ratio,
$\overline m=2.35m_H$ is the mean particle mass of gas, 
and $M_{\text{s}}=1M_{\odot}$ is the mass of the central star.
We use the isothermal sound speed $\cs$ and the mid-plane gas density $\rhog$ when estimating timescales.
If we choose $\Sigma_0=1.7\times10^3\rm g\ cm^{-2}$, $\cso=1.0\times10^5\rm cm\ sec^{-1}$,
$\Omega_0=2.0\times10^{-7} \rm sec^{-1}$, $R_0=1$ AU, $\fsd=0.01$, $q=3/2$, and $p=1/4$,
the disk profile is similar to the minimum mass solar nebula (MMSN, \cite{hay81}).

\subsection{Collision timescale}

Collision time scale is estimated as
\begin{equation}
\tcol\sim (\nd \cdot \pi \rd^2 \cdot <\vdd>)^{-1}\ ,\  \label{eq:tcol_def}
\end{equation}
where $\nd$ and $\vdd$ is the number density of the dust grains and
the relative velocity between the dust grains, respectively.
The parenthetic quantity $<Q>$ represents the statistical average.

The dust number density is  expressed as 
\begin{equation}
\nd=\frac{\Sigmad}{\hd \md}, \label{eq:nd}
\end{equation}
where  $\hd$ is the scale height of the dust layer, $\md$ is the mass of the dust grains. 
We approximate that the mass distribution function of the dust is the delta function 
since it is necessary for the dust grains to collide with the similar scale grains so that the grains gain the angular momentum. 
Considering the equilibrium between turbulent diffusion and sedimentation \cite{bir10}, 
$\hd$ is obtained as 
\begin{equation}
\hd=\hg\cdot \left(\frac{\alpha}{\rm St}\frac{\rm 1+2St}{\rm 1+St^2}\right)^{1/2},
\label{eq:hd}
\end{equation}
where $\st \equiv \Omegak \ts$ is the stokes number ($\ts$ is the stopping time by drag force). %\st=1/\gd
We use alpha prescription $\nuturb=\alpha \cs \hg$ to describe the strength of the turbulence in the protoplanetary disk,
and assume $\alpha < $ St to avoid the situation $\hg < \hd$.
At the Stokes drag law regime, the Stokes number is written as
\begin{equation}
{\rm St}=\frac{\md\Omegak}{6\pi \rd \rhog \nu}=\frac{2\rhoint\rd^2\Omegak}{3 \rhog \cs \mfp}
=\frac{2\sigmamol\rhoint \Omega_0}{3 \overline m \cso} \rd^2 R_1^{p-3/2} \label{eq:st},
\end{equation}
where $\rhoint$ is the internal mass density of the dust grains, 
$\nu= \cs \mfp/3$ is the kinematic viscosity, 
$\mfp$ is the mean free path of the gas particles. 
The mean free path is estimated as $\mfp=\overline{m}/(\sigmamol\rhog)$,
where $\sigmamol$ is the cross section of collisions between H$_2$ molecules. 
We adopt $\rhoint \simeq 3\ \rm g\ cm^{-3}$, and $\sigmamol\simeq 2\times 10^{-15}\ \rm cm^2$. 
Equation (\ref{eq:st}) means that the Stokes number is independent of the normalization
coefficient of the surface density $\Sigma_0$,
which is canceled out due to the one in $\mfp$.

Since the gas was assumed to be in a turbulent state described by the alpha prescription, 
we set the mean relative velocity $<\vdd>=<\vdd>_t$, 
where $<\vdd>_t$ means the relative velocity between the grains in the turbulent gas. 
According to \cite{orm07},  $<\vdd>_t$ with similar scale grains can be represented as 
\begin{equation} 
<\vdd>_t=\cs\left(\frac{\alpha {\rm St}}{\sqrt{1+\frac{1}{4}\rm St^2(1+St)^2}}\right)^{1/2},  \label{eq:wrelt}
\end{equation}
where we smoothly interpolate the two limiting solutions of $\st \gg1$ and $\st \ll1$.
This expression is valid when the stopping time is larger than the turn-over time of
the Kolmogorov-scale eddy.
The minimum size of the grain satisfying this condition is on the order of
sub-mm for MMSN at 1 AU,
so that we focus on the grain larger than $\sim$ 1 mm in what follows.

Now we can express $\tcol$ as the function of $\rd$ and $R_1$ by using Equations (\ref{eq:tcol_def}) to (\ref{eq:wrelt}) as
\begin{equation} 
\tcol=\frac{4\sqrt{2}\rhoint}{3\Omega_0\Sigma_0\fsd}\rd R_1^{q+3/2}f({\rm St}), \label{eq:tcol0}
\end{equation} 
where
\begin{equation}
f({\st})=\frac {1} {\rm St} \left(\frac{1+2\st}{1+\st^2}\right)^{1/2}\left(\rm 1+\frac{1}{4}\st^2(1+\st)^2\right)^{1/4}.
 \label{eq:fst}
\end{equation}
We note that when $\alpha <$ St, $\tcol$ is independent of $\alpha$
because the effect increasing $<\vdd>_t$ balances the one decreasing $\nd$. 
For the case with $\st\ll 1$, the collision time scale $\tcol \propto {\rm St}^{-1}\rd^1 R_1^{q+3/2} \propto \rd^{-1}R_1^{3+q-p}$,
while $\tcol\propto {\rm St}^{-1/2}\rd^1 R_1^{q+3/2} \propto \rd^0R_1^{2.25+q-0.5p}$ for the case with $\st \gg 1$. 
If we adopt the same parameters as MMSN,
collision time is
\begin{equation}
\tcol=1.7\times10^6 f({\rm St})R_1^3\rdone\ \rm sec, \label{eq:tcol} 
\end{equation}
where $\rdone=\rd/({\rm 1cm})$.

\subsection{Spin-down timescale}

In the case of the Stokes law, the angular momentum conservation around the spin axis of a spherical grain is given as
\begin{equation}
\id\frac{d\omegad}{dt}=-8\pi\rhog\nu \rd^3\omegad, \label{rot_eq}
\end{equation}
where $\id$ is the moment of inertia of the grain. 
The torque acting onto a spherical body by viscous fluid is given in \cite{rub61,tak74}.
From this equation, the $\tsd$ is estimated as
\begin{equation}
\tsd=\frac{\id }{8\pi \rhog \nu \rd^3}=\frac{\rhoint\rd^2}{5\rhog \cs\mfp}
 =\frac{\sigmamol\rhoint}{5\overline m \cso} \rd^2 R_1^p. \label{eq:tsd0}
\end{equation}
At the second equation, we assume a spherical and uniform density grain whose moment of inertia is represented $\id=2\md \rd ^2/5$. 
The spin-down time becomes longer as the dust grain becomes larger.
Here we note that the spin-down time is independent of $\Sigma_0$
by the same reason as the Stokes number (see Equation \ref{eq:st}).
For MMSN, $\tsd$ is estimated as 
\begin{equation}
\tsd=3.0\times10^3 \rdone^2 R_1^{1/4} {\rm sec}. \label{eq:tsd} 
\end{equation}

\subsection{Comparison of timescales}

Now, we can obtain the size of dust grains that are able to keep spinning. 
We estimate these time scales just in the Stokes law regime, 
because the lift force in other regime is uncertain. 
There are two necessary conditions to realize the Stokes law. 
One is that the gas can be regarded as a continuum medium, which is expressed as $\rd\gtrsim 9\mfp/4$. 
The other is that the flow around the dust grains is laminar, which is represented as  Re$=2 u \rd/\nu\lesssim 20$ \cite{shi92}, 
where $u$ is the relative velocity of the dust to the gas. 
Here, we should actually include the effect of turbulence into the expression of
$u$ as in \cite{orm07} so that
the physical situation is consistent with that of Equation (\ref{eq:wrelt}).
However, taking into account the effect causes complicated equations.
Thus, as a first-step study, we assume that $u$ is equal to the relative velocity
between the orbital velocity of the gas and the Keplerian velocity,
i.e., $u=\eta\vk$, where
\begin{equation}
\eta\equiv \frac{2p+2q+3}{4} \left(\frac{\cs}{\vk}\right)^2,
\end{equation}
which is given in \cite{ada76}.
By these conditions, we find that our estimation is valid in the range,
\begin{equation}
\begin{split}
\rdmin & \lesssim \rd \lesssim \rdmax, \text{ where} \\
&\rdmin \equiv \frac{9\overline m \cso}{2\sigmamol\Sigma_0\Omega_0}R_1^{q-p+3/2}, 
\text{ and}\\
&\rdmax \equiv \frac{40\sqrt{2\pi}\overline m R_0}{3(2p+2q+3)\sigmamol\Sigma_0} R_1^{q+1}.
\label{eq:dust_radii0}
\end{split}
\end{equation}
For MMSN, this condition is simply written as 
\begin{equation}
3.2R_1^{11/4}\lesssim \rdone\lesssim 89 R_1^{5/2}. \label{eq:dust_radii}
\end{equation}

\begin{figure}[t!]
\begin{center}
\includegraphics[width=8cm]{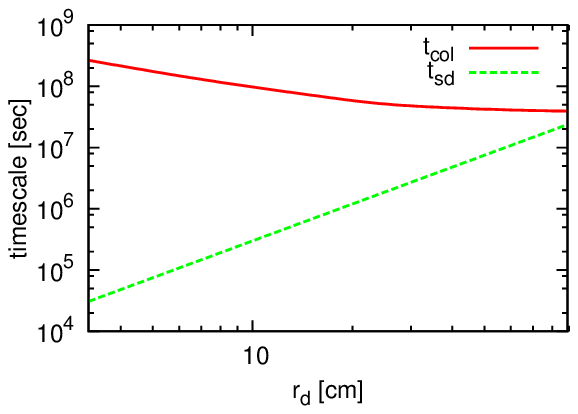}

\caption{$\rd$ dependence of $\tcol$ (solid lines) and $\tsd$(dashed lines) at $R_1 = 1$.
$\tcol$ is always longer than $\tsd$ in the Stokes regime.}

\label{fig:timescales}
\end{center}
\end{figure}

\begin{figure}[t!]
\begin{center}
\includegraphics[width=8cm]{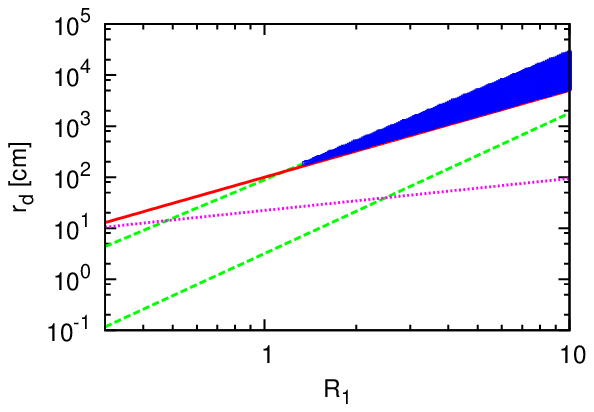}

\caption{Parameter space where the dust grains can keep spinning in the $\rd-R_1$ plane. 
The Stokes regime is between the two dashed green lines. 
The condition $\tcol < \tsd$ is satisfied above the solid red line. 
The grains may keep spinning in the blue region. The dotted magenta line
represents the line where the condition St$ = 1$ is satisfied.}

\label{fig:sustainability}
\end{center}
\end{figure}

Figure \ref{fig:timescales} shows that the two timescales $\tcol$ (solid lines) and $\tsd$ (dashed lines) at $R_1=1$ for MMSN. 
We plotted $\tcol$ and $\tsd$ in the range, which satisfies the condition (\ref{eq:dust_radii}). 
From Figure \ref{fig:timescales}, 
we can see that $\tcol$ is larger than $\tsd$,
%though the two timescales are comparable when the dust grains are large, $\rdone\sim 100$. 
so that the spin of the dust would stop at $R_1=1$.
The difference of two timescales are smaller as the dust grains are larger.
Equations (\ref{eq:tcol}) and (\ref{eq:tsd}) shows that large grains are likely to satisfy the condition $\tcol < \tsd$. 
From Equation (\ref{eq:dust_radii}), the Stokes regime can be adopted for the larger dust grains at the outer region of the disk. 
Thus, we expect that the condition $\tcol < \tsd$ is satisfied at the outer region $R_1 > 1$. 
Figure \ref{fig:sustainability} shows the parameter space where the dust grains keep spinning in $R_1-\rd$ plane for MMSN. 
The Stokes regime is realized between the dashed green lines. 
The condition $\tcol < \tsd$ is satisfied above the solid red line. 
In the blue region, the condition is satisfied with the Stokes regime. 
There are dust grains that keep spinning with the Stokes regime in $R_1 \gtrsim 1.3$.
The grains that can keep spinning have the size $\rd\sim\rdmax$. 
The dotted magenta line shows the dust radius when St $=1$, which is used in Section
\ref{dis}.

\section{Relative velocity between the dust grains}
\label{velo}
In this section, we investigate whether the mean relative
velocity is comparable to or greater than the gas velocity in the Kepler
rotational frame.
%, $\eta v_{\text{K}}$, where $v_{\text{K}}$ is the Kepler velocity.
Since this gas velocity is comparable to the typical relative velocity
between a large grain and a small one compared to one-meter-sized dust,
we take it as a reference value.
First, we derive the equation of motion for a dust grain assuming
that it moves at a terminal velocity.
Next we estimate the mean relative velocity by assuming
the isotropic distribution for the spin angular momentum.

Here, for simplicity, we assume that the dust grains move on
the mid-plane of the disk, which means the $z$-component of
the lift force is assumed to be zero, where $z$-axis is taken as
the disk axis, and we adopt below the cylindrical coordinate.
Since the direction of the spin angular momentum can be taken arbitrarily,
the lift force can show the $z$-component.
Nevertheless, we neglect the $z$-component of the velocity
to simplify the calculation below.

For the preparation to derive the equation of motion,
we express a projected vector of the lift force on the mid-plane in terms of
the direction of the spin angular momentum of a dust grain.
Since the direction of the lift force is perpendicular
to the spin angular momentum and the velocity of the grain with
respect to the gas, so that
\begin{equation}
\vec{F}_{\text{L}} = A \vec{\omega}_{\text{d}} \times \vec{u},
\end{equation}
where the coefficient satisfies
$A= \frac{\pi \rho_{\text{g}}r_{\text{d}}^3}{m_{\text{d}}}$
(see Section \ref{intro}).
Since $z$-component of $\vec{u}$ is zero, the lift force vector projected on
the mid-plane $\vec{F}_{\text{L,mid}}$ is expressed as,
\begin{equation}
\vec{F}_{\text{L,mid}} =
\vec{F}_{\text{L}} \cdot \begin{pmatrix} \vec{e}_{r} \\ \vec{e}_{\theta} \end{pmatrix}
= A \omega_{\text{d}}\mu \begin{pmatrix} -u_{\theta} \\ u_{r} \end{pmatrix}
\equiv F_{\text{L}} \begin{pmatrix} -u_{\theta}/u \\ u_{r}/u \end{pmatrix},
\label{flmu}
\end{equation}
where $\vec{e}$ with a subscript and $\mu$ represent
a unit vector in the direction of the subscript and
the cosine of the angle between $\vec{\omega}_{\text{d}}$ and
the $z$-axis, respectively.
We note here that the $F_{\text{L}}$ do not depend on the azimuth angle
of the spin angular momentum.

Next, we derive and reduce the equation of motion of the dust grain.
Now the forces exerted on the dust grain are
the gravitational force of the central star,
the drag force and the lift force, so the equation of motion is expressed as,
\begin{equation}
\frac{d\vec{v}}{dt} = - \frac{GM_{\odot}}{r^2} \vec{e}_{r}
- \vec{F}_{\text{D}} + \vec{F}_{\text{L,mid}},
\label{eomvec}
\end{equation}
where we assume that the mass of the central star is the same as the solar one.
As the first step of the reduction of Equation (\ref{eomvec}),
we divide it into two equations for $r$ and $\theta$ components.
Since the velocity of the disk gas does not have the radial component, 
the components of the velocity of the dust grain are represented as
$(u_{r}, u_{\theta}) = (v_{r}, v_{\theta}-r\Omega_{\text{g}})$,
where $\Omega_{\text{g}}$ is the orbital angular velocity of the disk gas
around the central star.
Thus, Equation (\ref{eomvec}) is expressed as,
\begin{gather}
\frac{dv_{r}}{dt} - \frac{v_{\theta}^2}{r} = - \frac{GM_{\odot}}{r^2} 
- F_{\text{D}} \frac{v_{r}}{u} - F_{\text{L}} \frac{v_{\theta}-r\Omega_{\text{g}}}{u}, \\
\frac{dv_{\theta}}{dt} + \frac{v_{r}v_{\theta}}{r} = 
- F_{\text{D}} \frac{v_{\theta}-r\Omega_{\text{g}}}{u} + F_{\text{L}} \frac{v_{r}}{u}.
\label{eomcom}
\end{gather}
As the second step, we transform this into the coordinate
rotating at the angular velocity of the Kepler rotation, that is,
$v_{\theta} = v_{\text{K}} + v_{\theta}^{\prime}$.
As the third step, we assume that the motion of the dust grain is stationary,
and that $|v_{r}|, |v_{\theta}^{\prime}| \ll v_{\text{K}}$.

This stationary assumption may be invalid taking into account
the timescales discussed in Section \ref{sustainability}.
The stopping time $\ts$ is represented as,
\begin{equation}
\ts=\frac{\md}{6\pi \rd \rhog \nu}
\sim 10^{4}\ \rdone^2R_1^{1/4}
\sim 4\ \tsd. \label{eq:ts}
\end{equation}
This means that the dust grain stops spinning before moving at
the terminal velocity independently of the dust size and the distance
from the central star.
Thus, as long as the lift force is exerted on the grain,
the motion of the grain cannot reach a steady state.
Alternatively, the grain motion is considered to 
be determined by the merger of the parent grains (or scattering by
the other grain).
Nevertheless, we assume that the grain motion reaches the steady state,
as the first stage of this kind of work.

Finally, we nondimensionalize {the variables} as
\begin{equation}
x = \frac{v_{r}}{\eta v_{\text{K}}},\ 
y = \frac{v_{\theta}^{\prime}}{\eta v_{\text{K}}},\ 
\gd = \frac{F_{\text{D}}}{u\Omegak},\
\gl = \frac{F_{\text{L}}}{u\Omegak},
\label{nondime}
\end{equation}
where $\eta$ is the constant satisfying the equation
$r\Omega_{\text{g}} = v_{\text{K}} (1-\eta)$, and
$\Omegak$ is the angular velocity of the Kepler motion.
Thus, we obtain two algebraic equations
\begin{gather}
2y = \gd x + \gl (y+1), \\
\frac{1}{2}x = - \gd (y+1) + \gl x.
\end{gather}
These equations represent the balance between Coriolis, drag and lift forces.

\begin{figure}[t!]
\begin{center}
\includegraphics[height=8cm,angle=270,clip]{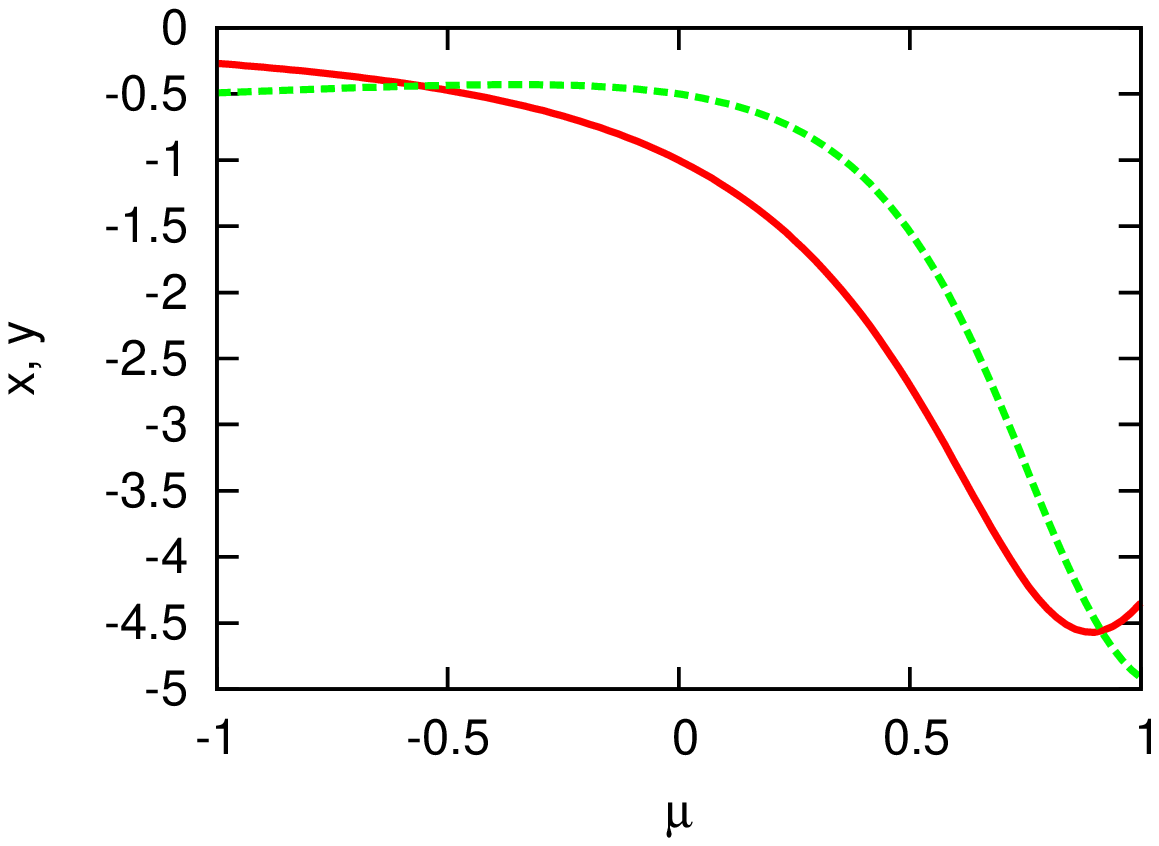}
\caption{Dependence of the grain velocity in the radial direction $x$
(the solid line) and in the azimuthal direction $y$ (the dashed line)
on the cosine of the angle between
the spin angular momentum and $z$-axis $\mu$.
We take the parameters as $\gd = 1$ and $\rld = 1$.}
\label{xyvsmu}
\end{center}
\end{figure}

The solution of the equations is
\begin{gather}
x = \frac{-2\gd}{\gd^2 + (\gl - 2)(\gl - 1/2)} \label{xmu}, \\
y = -1 + \frac{-2(\gl - 1/2)}{\gd^2 + (\gl - 2)(\gl - 1/2)}.\label{ymu}
\end{gather}
Here, Equations (\ref{flmu}) and (\ref{nondime}) lead to
$\gl = \glm \mu$, where $\glm \equiv \frac{A\omegad}{\Omega_{\text{g}}}$,
and we introduce lift-drag ratio $\rld \equiv \glm / \gd$ to obtain
$\gl = \gd \rld \mu$.
Thus, $x$ and $y$ are expressed as functions of $\mu$, $\gd$ and $\rld$.
We show $x(\mu)$ and $y(\mu)$ in Figure \ref{xyvsmu}.
Here we assume $\gd = 1$ and $\rld = 1$ as a trial value.
%, which correspond to
%the situation that the stopping time is equal to the Kepler rotation time 
%and that the spinning bodies move in a viscous gas slowly flowing, respectively.
Since $x = -1$ when we neglect the lift force, the curve of $x(\mu)$
shows the radial velocity of the dust grain can be a third or
four times compared to that without the lift force.
On the other hand, when $\mu < 0$,
$y$ is almost constant and comparable to that without the lift force.
When $\mu$ is larger than 0.5, $y$ is smaller than $-1$,
which means that the dust grain orbits more slowly than the gas.
In addition, we see that $x$ and $y$ decrease as $\mu$ is close to unity.
Therefore, the absolute value of the velocity tends to increase
as $\mu$ increases.

Next, we calculate the average and dispersion of the velocity of the dust grain
on the disk mid-plane, assuming that the spin angular momentum
is isotropic, which is just for simplicity.
Thus, the direction distribution satisfies $f(\Omega) = \frac{1}{4\pi}$,
which is equivalent to $f(\mu) = \frac{1}{2}$, where $\Omega$ is
a solid angle parameter.
The average and dispersion of $x$ are calculated by performing
the integration below.
\begin{equation}
\begin{split}
<x>\ & = \int xf(x)dx\\
    & = \frac{1}{2}\int^1_{-1}x(\mu)d\mu,\\
<x^2>\ & = \int x^2f(x)dx\\
    & = \frac{1}{2}\int^1_{-1}x^2(\mu)d\mu,
\end{split}
\label{xave}
\end{equation}
where we transform the integration variable into $\mu$.
We also can derive the same expression for $y$,
\begin{equation}
\begin{split}
<y>\ & = \frac{1}{2}\int^1_{-1}y(\mu)d\mu, \\
<y^2>\ & = \frac{1}{2}\int^1_{-1}y^2(\mu)d\mu.
\label{yave}
\end{split}
\end{equation}
Taking $\gd = 1, \rld = 1$, we obtain the approximate value
of the average and standard deviation of the velocity,
\begin{align}
<x>\ \simeq -1.4  \label{xvlu},\\
<y>\ \simeq -1.2 \label{yvlu},\\
\sqrt{<x^2>}\ \simeq 1.0  \label{x2vlu},\\
\sqrt{<y^2>}\ \simeq 0.6 \label{y2vlu},\\
<\wrel> \equiv \sqrt{<x^2>+<y^2>} \simeq 1.2. \label{vvlu}
\end{align}
Equation (\ref{xvlu}) means that the dust grains averagely fall down
to the star faster than without the lift force.
On the other hand, Equation (\ref{yvlu}) means that they averagely orbit 
at almost the same velocity as the gas.
We note that the standard deviation of the velocity represents
average of the relative velocity between the grains.
Therefore, Equation (\ref{vvlu}) means that the average relative
velocity exceeds the relative velocity between the gas and
Kepler velocity, so that the collision rate is affected by
the lift force when $\gd = 1$ and $\rld = 1$.
% here, do we mention that <vdd> < <wrel> when St~1, alpha~10^-3 at R0=1.

\begin{figure}[t!]
\begin{center}
\includegraphics[height=8cm,angle=270,clip]{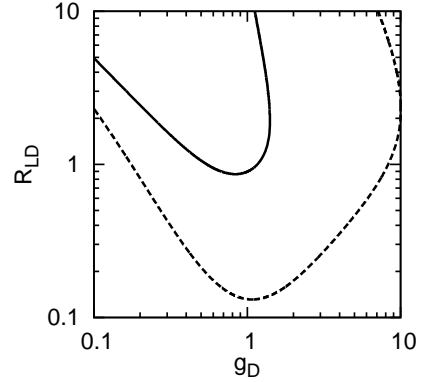}
\caption{Contours of the averaged relative velocity of the grains are
shown on the $\gd$-$\rld$ plane. The solid and dashed lines represent
$<\wrel> =$ 1 and 0.1, respectively.}
\label{sigma2}
\end{center}
\end{figure}

We finally calculate the averaged relative velocity on the disk mid-plane
$<\wrel>$ for arbitrary values of $\gd$ and $\rld$.
Figure \ref{sigma2} shows the contour lines of $<\wrel> =$1 (the solid line) and
0.1 (the dashed line) on the $\gd$-$\rld$ plane.
We see that the relative velocity is large when $\gd$ is small and when
$\rld$ is large, which corresponds the situation that
the lift force is efficiently exerted on the grains.
The important fact is that there exist a region satisfying
$<\wrel> > 1$, where the lift force non-negligibly affects the dynamics
of the system of grains, compared to the case without the force.

\section{Discussion}
\label{dis}
\subsection{Lift-drag ratio}

In this subsection, we estimate the lift-drag ratio $\rld$
to investigate how efficiently the lift force affects the motion of the dust. 
In the Stokes regime, the lift coefficient $\cl$ and drag coefficient $\cd$ is
represented as
\begin{eqnarray}
\cd&=&\frac{4 \cs \mfp}{\rd u}, \label{C_D}\\
\cl&=&\frac{2\md\fl}{\pi \rd^2 \rhog u^2}=\frac{2\rd\omegad}{u}\sin\theta \label{C_L}. 
\end{eqnarray}

The lift coefficient depends on the spin angular velocity $\omegad$ of the dust.
Here, we estimate $\omegad$ induced by the collisions of the grains. 
When two grains with the same mass $\md$ collide with the impact parameter $b$,
the angular momentum around the center of mass is represented by 
\begin{equation}
 L= \frac {b \vdd \md} 2 .
\end{equation}
Given the weight by a cross section, 
we derive the averaged angular velocity $\sqrt{<L^2>}$ as 
\begin{equation}
\sqrt{<L^2>}=\left(\frac{\int^{2\rd}_0 L^2 2\pi b db}{\int^{2\rd}_0 2\pi b db}\right)^{1/2}
=\frac{1}{\sqrt{2}}<\vdd> \rd \md. 
\end{equation}
If we assume that the grain obtains the mass $2\md$  and this angular momentum after the collision, 
the resultant angular velocity $\omegad$ is represented as 
\begin{equation}
\omegad=\frac{5\sqrt{2}}{4}\frac{<\vdd>}{\rd}  \label{omega_d}.
\end{equation}

Using Equation (\ref{C_D}),  (\ref{C_L}), and (\ref{omega_d}), 
we can reduce the  lift-drag ratio $\rld$ as
\begin{equation}
\rld=\frac{\cl}{\cd}=\frac{5\sqrt{2} \rd <\vdd>}{8 \cs\mfp}.
\label{eq:rld0}
\end{equation}
Futhermore, we adopt $<\vdd>=<\vdd>_t$, so that this expression is nearly
independent of the dust radius $\rd$ for $\st > 1$. 
For $\st \gg 1$, the lift-drag ratio approaches asymptotically
to a maximum,% for $\rd$, %which is represented as
\begin{equation}
\rld \simeq 20\alpha^{1/2}R_1^{-17/8},
\label{ldratio}
\end{equation}
where we take the MMSN disk parameters. 

We see from Figure \ref{fig:sustainability} that when $\rd \sim 10^2$ cm
at $R_1\sim 1$, which corresponds to $\gd = \st^{-1} \sim 0.1$, then
the conditions that $\tcol \sim \tsd$ and that the drag
force is represented with the Stokes law are marginally satisfied.
In this case, we find $\rld \sim 2.0$ when $\alpha = 0.01$,
using Equation (\ref{ldratio}).
Thus, from Figure \ref{sigma2} we obtain $<\wrel> \sim 0.1$,
which means that the averaged
relative velocity due to the lift force is a tenth of $\eta v_{\text{K}}$
for the one-meter sized dust at 1 AU from the central star.

\subsection{Dependence of the relative velocity on disk parameters}
The situation stated in the previous subsection can be qualitatively
or quantitatively changed by adopting disk parameters different from MMSN.
If we take larger $\fsd$ than 0.01, as proposed in \cite{has14,sek98},
the averaged relative velocity $<\wrel>$ can be $\sim 1$.
The dust to gas ratio $\fsd$ is included just in the expression of $\tcol$ (Equation
\ref{eq:tcol0}), so that larger $\fsd$ means smaller $\tcol$ and thereby smaller
$\rd$ satisfying $\tcol = \tsd$.
This implies that the red line in Figure \ref{fig:sustainability} moves down
and that the blue region expands inside.
When $\fsd \sim 0.1$, the blue region includes the dotted magenta line, which shows
the grain size satisfying St $= 1$, at $R_1 \sim 0.5$.
Since $\rld \sim 1$ when $\alpha = 10^{-4}$, we find from Figure \ref{sigma2}
that the relative velocity $<\wrel> \sim 1$.
In this case, the relative velocity due to the lift force $\eta \vk <\wrel>
\sim 60$ m s$^{-1}$, where we note that $\eta \vk$ does not depend on
$R_1$ when $p=1/4$, exceeds that due to the turbulence $<\vdd>_t \sim 8$
m s$^{-1}$, which is computed with Equation (\ref{eq:wrelt}).
Therefore, the lift force is expected to efficiently affect
the growth rate of the grain whose
size is 10 cm at $R_1 \sim 0.5$ if there is a grain-concentrated part.

Recently, the steeper density profile, which is denser at 1 AU, has been proposed
\cite{des07}.
When the density profile is steeper, a larger $q$ is adopted.
If $q$ is larger, the radial profile of the minimum and maximum sizes of the grains is
steeper (Equation \ref{eq:dust_radii0} and dashed green lines
in Figure \ref{fig:sustainability}).
On the other hand, we can find that the red line in Figure \ref{fig:sustainability}
does not change as much as the green lines, by comparing Equation (\ref{eq:tcol0}) to 
(\ref{eq:tsd0}).
Thus, the blue region in Figure \ref{fig:sustainability} slightly shifts to inner region.
Therefore, we expect that when the density profile is steeper with the normalization
coefficient fixed, the innermost radius where the lift force continues to be
exerted on the grains slightly decreases to get close to $R_1 = 1$.

Suppose that the disk has larger surface density. 
Spin-down time (Equation \ref{eq:tsd0}) and Stokes number (Equation \ref{eq:st})
are independent of the surface density $\Sigma_0$. 
On the other hand, collision time is inversely proportional to $\Sigma_0$
(Equation \ref{eq:tcol0}),
so that $\rd$ satisfying $\tcol = \tsd$ is
proportional to $\Sigma_0^{-1/2}$ for St $\gg 1$.
In contrast, $\rdmin$ and $\rdmax$ in Equation (\ref{eq:dust_radii0}) are inversely
proportional to $\Sigma_0$.
Therefore, the blue region moves outside and the mininum grain size in the blue
region is nearly-unchanged.
The ratio $\rld$ becomes smaller with decreasing $R_1$, so that
the relative velocity due to the lift force is smaller than that for the fiducial
surface density.

\subsection{Other effects for the model refinement}
% neglected process
We can refine the model in this paper by taking into account realistic
porosity and shape of the dust grain \cite{suy08}.
If the grain is fluffy, $\rhoint$ is smaller than the value we use in Section \ref{sustainability}.
Equations (\ref{eq:st}, \ref{eq:tcol}, \ref{eq:fst}) lead to the dependence of $t_{\text{col}}$
on $\rhoint$,
\begin{equation}
\tcol \propto
   \begin{cases}
    \rhoint^{\frac{1}{2}} & \text{ for } \st \gg 1 \\
    \rhoint^0 & \text{ for } \st \ll 1,
   \end{cases}
\end{equation}
while $\tsd \propto \rhoint^1$.
Thus, when $\rhoint$ decreases, $\tsd$ decreases more rapidly than
$\tcol$, which means that the lift force is exerted on the grain
in shorter time.
If the grain is lumpy, the coefficient of lift would become as large
as a baseball or a golf ball.

It is also worth taking into account realistic collision processes
between the grains such as simple scattering (bouncing), minor merger and destruction.
Through the simple scattering, the grain may gain the spin angular
momentum by means of the surface friction, where the maximum surface velocity
of the scattered grains is $\vdd$, which is less than the mean surface velocity in
the case of the major merger (Equation \ref{omega_d}).
In addition, in the case of a minor merger, which is realized when
we consider the size distribution of grains \cite{win12},
the grains obtain less spin angular momentum compared to the case of
the major merger.
Thus, we expect less mean relative velocity when the grains undergo
the scattering and minor merger.
On the other hand, if the destruction (fragmentation) occurs when the grains
collide, the grains may gain larger spin angular momentum.
The experiments show that agglomerates (cm to dm size) are divided into many fragments
that are of mm to cm size through the low-velocity collision \cite{dec13,sch12}.
If these fragments have a large spin angular momentum, they can be 
sufficiently affected by the lift force.
How much spin angular momentum grains gain depends on many parameters, so that
we defer it to the future work.

% effects of z-direction lift force
The z-component of the lift force, which is omitted in this paper,
may affect the resultant relative velocity of the dust grain.
The equation of motion in the z-direction includes
the gravitational force by the central star, the drag force and
the lift force.
Although the steady state cannot be realized, because the gravitational
force depends on the altitude from the disk mid-plane, the dust grains
should gain a momentum in the z-direction.
This causes an increase in the absolute value of the velocity,
which may result in an increase in relative velocity.

\section{Conclusion}
\label{sum}
In this paper, we investigate the effects of the lift force on the dust
grains in the protoplanetary disk from two perspectives.
We first investigate whether the lift force is kept exerted on
the grains or not.
We assume the grains are in the minimum mass solar nebula
where the turbulence is developed.
We estimate the collision timescale and the spin-down timescale and
find that the grain keeps spinning due to
the collision with the other grains if the radius of the grain is 
larger than 100 cm at $\gtrsim$ 1 AU from the central star.

We next calculate the mean relative velocity between the grains
caused by the lift force.
The grains obtain spin angular momenta with various directions
by the collision between themselves, so that the lift forces exerted on them
have the various directions. % is  in the disk gas.
Thus, the relative velocity yields between the grains.
We assume that the grains are 
in the steady state and that the distribution of their spin momenta
shows the isotropy.
We show that the mean relative velocity is comparable to
the gas velocity at the Kepler rotational frame,
when $\fl \gtrsim \fdr$ and $\ts \sim 1/\Omegak$, where $\fl, \fdr,
\ts \text{ and } \Omegak$ are the lift force, the drag force, the stopping
time of the grains by the drag, and the Kepler angular velocity,
respectively.
This means that the lift force can sufficiently affect the collision rate,
which affects the growth rate of the grains, under the parameter set.

We also estimate the mean relative velocity when the grains keep
spinning by combining the above two results.
We obtain the result that for the minimun mass solar nebula
the mean relative velocity due to the lift force 
is smaller than the gas velocity at the Kepler rotational frame.
We discuss the mean relative velocity as being comparable to
the gas velocity
if the disk has grain-concentrated parts where
the dust-gas ratio is ten times larger than MMSN,
so the lift force may affect the collision rate in the parts.

%%%%%%%%%%%%%%%%%%%%%%%%%%%%%%%%%%%%%%%%%%%%%%
%%                                          %%
%% Backmatter begins here                   %%
%%                                          %%
%%%%%%%%%%%%%%%%%%%%%%%%%%%%%%%%%%%%%%%%%%%%%%

\begin{backmatter}

%\section*{List of abbreviations}
%MMSN: minimum mass solar nebula
%
%\section*{Competing interests}
%  The authors declare that they have no competing interests.
%
%\section*{Author's contributions}
%MSY carried out the estimation of the mean relative velocity of the grains,
%participated in the sequence alignment and drafted the manuscript.
%SSK carried out the estimation of the condition under which the lift force
%is kept exerted on the grains,
%participated in the sequence alignment and drafted the manuscript.
%All authors read and approved the final manuscript.
%
\section*{Acknowledgements}
M. S. Y. and S. S. K. would thank T. Tsuribe, S. Inutsuka and S. Okuzumi
for useful discussions and comments and
acknowledge the anonymous referees for useful comments.
S. S. K. is supported by a Grant-in-Aid for JSPS Research Fellowships
for Young Scientists (A2517840).

%%%%%%%%%%%%%%%%%%%%%%%%%%%%%%%%%%%%%%%%%%%%%%%%%%%%%%%%%%%%%
%%                  The Bibliography                       %%
%%                                                         %%
%%  Bmc_mathpys.bst  will be used to                       %%
%%  create a .BBL file for submission.                     %%
%%  After submission of the .TEX file,                     %%
%%  you will be prompted to submit your .BBL file.         %%
%%                                                         %%
%%                                                         %%
%%  Note that the displayed Bibliography will not          %%
%%  necessarily be rendered by Latex exactly as specified  %%
%%  in the online Instructions for Authors.                %%
%%                                                         %%
%%%%%%%%%%%%%%%%%%%%%%%%%%%%%%%%%%%%%%%%%%%%%%%%%%%%%%%%%%%%%

% if your bibliography is in bibtex format, use those commands:
\bibliographystyle{bmc-mathphys} % Style BST file
\bibliography{bmc_article}      % Bibliography file (usually '*.bib' )

\end{backmatter}
\end{document}